\documentclass[10pt,showpacs,twocolumn,nofootinbib,aps,prd]{revtex4-1}
\usepackage{makecell}
\usepackage{amsmath,amssymb,bm,mathrsfs}
\usepackage{mathtools,mathptmx,array,slashed}
\usepackage{graphicx,graphics,subfigure}
\usepackage{ctable,multirow}
\usepackage{tabularx}
\usepackage[colorlinks=true,linkcolor=blue,citecolor=blue,urlcolor=blue]{hyperref}
\usepackage{color}

\DeclareSymbolFont{symbols} {OMS}{cmsy}{m}{n}

\def\be{\begin{equation}}
\def\ee{\end{equation}}
\def\bea{\begin{eqnarray}}
\def\eea{\end{eqnarray}}
\def\ba{\begin{aligned}}
\def\ea{\end{aligned}}
\def\nn{\nonumber}
\def\p{\partial}
\newcommand{\revise}[1]{{\color{red}#1}}

\begin{document}

%\begin{CJK*}{GBK}{song}

\title{Dimensional structure of thermodynamic topology in ultraspinning Kerr-AdS black holes}

%\author{Min Tian (田敏)}
\author{Min Tian}

%\author{Ying Chen (陈颖)}
\author{Ying Chen}

%\author{Di Wu (吴迪)}
\author{Di Wu}
\email{Corresponding author: wdcwnu@163.com}
%https://orcid.org/0000-0002-2509-6729

\affiliation{School of Physics and Astronomy, China West Normal University, Nanchong, Sichuan 637002, People's Republic of China}

\date{\today}

\begin{abstract}
In this paper, we apply the thermodynamic topology framework to ultraspinning Kerr-AdS black holes in arbitrary spacetime dimensions. By constructing the off-shell Helmholtz free energy and the associated vector field, black hole states are characterized as topological defects, and their phase structures are described through zero points, winding numbers, and asymptotic thermodynamic behavior. Analyses of the four- and five-dimensional cases highlight the differences between even- and odd-dimensional configurations, while the endpoint behavior of the inverse-temperature curve, together with representative higher-dimensional cases, supports the absence of additional topological classes or subclasses. We find that only two thermodynamic topological structures appear: the standard class $W^{1+}$ for most configurations, and the distinct subclass $\tilde{W}^{1+}$ for odd-dimensional black holes with maximal rotations. These results support a unified classification scheme valid across dimensions for ultraspinning Kerr-AdS black holes.
\end{abstract}

\maketitle

%\end{CJK*}

%%%%%%%%%%%%%%%%%%%%%%
\section{Introduction}
%%%%%%%%%%%%%%%%%%%%%%
In recent years, a new class of ultraspinning anti-de Sitter (AdS) black holes \cite{PRL115-031101,JHEP0114127,PRD89-084007}, obtained by boosting one rotational angular velocity to the speed of light, has attracted considerable attention. These black holes preserve a finite horizon area while exhibiting a noncompact horizon topology, manifested as two punctures at the north and south poles of the spherical horizon. They can violate the reverse isoperimetric inequality (RII)~\cite{PRD84-024037,PRD87-104017}, indicating that the Schwarzschild-AdS black hole represents the maximal entropy configuration. Since the entropy of ultraspinning black holes exceeds this bound, they are widely recognized as superentropic black holes.

As shown in Ref.~\cite{PRL115-031101}, such ultraspinning solutions can be generated via a straightforward ultraspinning limit applied to ordinary rotating AdS black holes. This procedure involves rewriting the metric in a rotating frame at infinity, boosting one rotational angular velocity to the speed of light, and compactifying the associated azimuthal direction. Following this method, a variety of ultraspinning black hole solutions have been constructed from known rotating AdS spacetimes \cite{JHEP0615096,PRD95-046002,1702.03448,
JHEP0118042,PRD102-044007,PRD103-044014,JHEP1121031}. More recently, an alternative construction has been proposed, in which superentropic black holes emerge through the introduction of a conical deficit in rotating AdS geometries \cite{JHEP0220195}. Extensive investigations have explored their physical properties, including thermodynamics \cite{PRL115-031101,JHEP0615096,PRD95-046002,1702.03448,JHEP0118042,MPLA35-2050098,PRD101-086006,
PRD101-024057,PLB807-135529,2601.22565}, horizon geometry \cite{PRD89-084007,JHEP0615096,
PRD95-046002,PRD102-044007,PRD103-044014,JHEP1121031}, Kerr/conformal field theory (CFT) correspondence \cite{PRD95-046002,1702.03448,JHEP0816148}, geodesic motion \cite{1912.03974}, null hypersurface caustics~\cite{CQG38-045018,PRD103-024053,PRD103-104020,PRD104-L121501}, black hole shadows \cite{PLB821-136619,EPJC82-619}, and related phenomena. Among these directions, thermodynamic properties play a central role in understanding ultraspinning black holes, naturally motivating a classification based on their thermodynamic phase structure.

Within this context, topology provides a systematic framework to classify black hole phases through global, quantitative invariants \cite{PRL129-191101,PRD110-L081501,PRD111-L061501,
PRD112-124024,2510.20164,EPJC85-1386}. By modeling solutions as topological defects in the thermodynamic parameter space, black holes are organized into well-defined topological classes. Initial studies identified three categories characterized by topological numbers \cite{PRL129-191101}, which were later generalized into four broader classes via asymptotic thermodynamic behavior analysis \cite{PRD110-L081501}. This classification framework was subsequently further developed to encompass five topological classes and five subclasses \cite{PRD111-L061501,PRD112-124024,2510.20164,EPJC85-1386}. \footnote{Additional representative examples are compiled in Refs. \cite{PRD107-024024,PRD107-064023,PRD107-084002,
EPJC83-365,EPJC83-589,PRD108-084041,JHEP0624213,PLB856-138919,PDU46-101617,EPJC84-1294,
CQG42-125007,PLB860-139163,PLB865-139482,EPJC85-828}, which summarize recent developments in this field. Beyond thermodynamic classification, this topological framework has also been applied to the study of light rings \cite{PRL119-251102,PRL124-181101,PRD102-064039,
PRD103-104031,PLB858-139052,PLB868-139742} and timelike circular orbits \cite{PRD107-064006,
JCAP0723049,PRD108-044077,CQG42-025020}, demonstrating its broad applicability.} It thus provides a natural theoretical foundation for the thermodynamic topological classification of ultraspinning black holes.

In this work, we employ the thermodynamic topological framework to study ultraspinning Kerr-AdS black holes in arbitrary dimensions and systematically classify their topological (sub)classes. Our analysis reveals a distinct pattern: in odd dimensions, black holes with the maximal number of rotation parameters form the subclass $\tilde{W}^{1+}$, whereas all other configurations, including odd-dimensional black holes with fewer rotations and all even-dimensional cases, belong to the class $W^{1+}$. The remainder of this paper is organized as follows. In Sec. \ref{II}, we provide a concise overview of the thermodynamic topological method. In Sec. \ref{III}, the four-dimensional and five-dimensional ultraspinning Kerr-AdS black holes are examined as representative cases to illustrate the behaviors of even- and odd-dimensional configurations. The generalization to higher dimensions ($d \ge 6$), which tests and further develops the same classification pattern, is presented in Sec. \ref{IIIC}. Finally, Sec. \ref{IV} summarizes our main findings and conclusions.

%%%%%%%%%%%%%%%%%%%%%%%%%%%%%%%%%%%%%%%%%%%%%%%%%%%%%%%%%%%%%%%%%%%%%%%%%%%%
\section{Overview of the thermodynamic topological framework}\label{II}
%%%%%%%%%%%%%%%%%%%%%%%%%%%%%%%%%%%%%%%%%%%%%%%%%%%%%%%%%%%%%%%%%%%%%%%%%%%%
We briefly review the thermodynamic topological approach in this section. The starting point is the generalized off-shell Helmholtz free energy \cite{PRD110-L081501}:
\be\label{FE}
\mathcal{F} = M - \frac{S}{\tau} \, ,
\ee
where $M$ and $S$ denote the black hole mass and Bekenstein-Hawking entropy, respectively, and $\tau$ represents the inverse temperature of a cavity enclosing the black hole, allowing an off-shell formulation. The on-shell condition is realized uniquely when $\tau = \beta = 1/T$, at which point $\mathcal{F}$ reduces to the conventional Helmholtz free energy $F = M - TS$ \cite{PRD15-2752,PRD33-2092,PRD105-084030,PRD106-106015}.

Following Ref.~\cite{PRD111-L061501}, a two-component vector field can be introduced via
\be
\phi = \big(\phi^{r_h}, \phi^\Theta \big) = \left(\frac{\partial \mathcal{F}}{\partial r_h}, -\frac{\cos\Theta}{\sin^2 \Theta}\right) \, ,
\ee
where $r_h$ is the event horizon radius and $\Theta\in(0,\pi)$ is an auxiliary variable. The endpoints $\Theta=0,\pi$ are excluded because $\phi^\Theta$ is singular there, while within this angular interval the second component vanishes uniquely at $\Theta=\pi/2$. The zero points of the vector field, associated with black hole solutions at inverse temperature $\tau$, are determined as follows. Since the second component vanishes only for $\Theta=\pi/2$, attention is restricted to the first component. Its vanishing condition is
\be
\phi^{r_h} = \frac{\p M}{\p S}\frac{\p S}{\p r_h} -\frac{1}{\tau}\frac{\p S}{\p r_h}
= \frac{\p S}{\p r_h}\Big(T -\frac{1}{\tau}\Big) = 0 \, ,
\ee
which directly leads to $\tau = 1/T$ as the defining relation for the solutions.

Using Duan's $\phi$-mapping theory \cite{SS9-1072,NPB514-705,PRD61-045004}, one can define a conserved topological current. Denoting $x^\nu = (\tau, r_h, \Theta)$ and the normalized vector field $n^a$ with components $n^r = \phi^{r_h}/||\phi||$ and $n^\Theta = \phi^{\Theta}/||\phi||$, the current reads
\be
j^\mu = \frac{1}{2\pi} \epsilon^{\mu\nu\rho} \epsilon_{ab} \partial_\nu n^a \partial_\rho n^b \, , \qquad \mu,\nu,\rho = 0, 1, 2,
\ee
which is identically conserved, $\partial_\mu j^\mu = 0$. Equivalently, it can be written as
\be
j^\mu = \delta^2(\phi) J^\mu\Big(\frac{\phi}{x}\Big),
\ee
indicating non-vanishing contributions only at zero points $\phi^a(x_i) = 0$. The total topological number in a region $\Sigma$ is
\be
W = \int_\Sigma j^0 \, d^2x = \sum_{i=1}^N \beta_i \eta_i = \sum_{i=1}^N w_i,
\ee
where $\beta_i$ and $\eta_i$ denote the Hopf index and Brouwer degree, respectively. This intrinsic winding number $w_i$ characterizes each zero point, independent of the enclosing curve.

Crucially, this formalism allows the assignment of stability: $w = +1$ for locally stable states and $w = -1$ for unstable ones. Hence, the global topological number $W$ provides a robust classification of black holes solely in terms of topological invariants.

%%%%%%%%%%%%%%%%%%%%%%%%%%%%%%%%%%%%%%%%%%%%%%%%%%%%%%%%%%%%%%%%%%%%%%%%%%%%%%%
\section{Ultraspinning Kerr-AdS black holes in arbitrary dimensions}\label{III}
%%%%%%%%%%%%%%%%%%%%%%%%%%%%%%%%%%%%%%%%%%%%%%%%%%%%%%%%%%%%%%%%%%%%%%%%%%%%%%%
In this section, we investigate the topological (sub)classes identified for ultraspinning Kerr-AdS black holes in all dimensions and provide a systematic analysis of their asymptotic thermodynamic behavior. In the generalized Boyer-Lindquist coordinates, the metric of the $d$-dimensional ultraspinning Kerr-AdS black holes is given by \cite{JHEP0615096}:
\be
ds^2 = d\gamma_s^2 +\frac{2m}{U}\omega_s^2 +\frac{Udr^2}{F -2m} +d\Omega_s^2 \, ,
\ee
where
\bea
d\gamma_s^2 &=& -\left[(\hat{W} +\mu_j^2)\rho^2 +\mu_j^2 l^2 \right]\frac{dt^2}{l^2} +\frac{2\rho^2 \mu_j^2}{l}dtd\varphi_j \nn \\
&&+\sum_{i\neq j}\frac{r^2 +a_i^2}{\Xi_i}\mu_i^2d\varphi_i^2 \, , \nn \\
U &=& r^\epsilon\left(\mu_j^2 +\sum_{i\neq j}\frac{\mu_i^2\rho^2}{r^2 +a_i^2} \right)\prod_{k \neq j}^N(r^2 +a_k^2) \, , \nn \\
\omega_s &=& (\hat{W} +\mu_j^2)dt -l\mu_j^2d\varphi_j -\sum_{i\neq j}\frac{a_i}{\Xi_i}\mu_i^2d\varphi_i \, , \nn \\
F &=& \frac{r^{\epsilon -2}\rho^4}{l^2}\prod_{i \neq j}^N(r^2 +a_i^2) \, , \nn \\
d\Omega_s^2 &=& \sum_{i \neq j}\frac{r^2 +a_i^2}{\Xi_i}d\mu_i^2 -2\frac{d\mu_j}{\mu_j}\left(\sum_{i \neq j}^{N +\epsilon}\frac{r^2 +a_i^2}{\Xi_i}\mu_id\mu_i \right) \nn \\
&&+\frac{d\mu_j^2}{\mu_j^2}(\rho^2\hat{W} +l^2\mu_j^2) \, , \nn
\eea
in which
\be
\hat{W} =  \sum_{i \neq j}\frac{\mu_i^2}{\Xi_i} \, , \quad \rho^2 = r^2 +l^2 \, , \quad \Xi_i = 1 -\frac{a_i^2}{l^2} ~~{\rm for}~~ i \neq j \, . \nn
\ee
Here, $m$ is the mass parameter, $a_i$ are the independent rotation parameters, and $l$ is the AdS radius. To treat spacetimes in both even and odd dimensions in a unified manner, we introduce the parameter $\epsilon$, defined as $\epsilon = 1$ for even dimensions and $\epsilon = 0$ for odd dimensions. The spacetime dimension $d$ can then be written as
\be
d = 2N +1 +\epsilon \, .
\ee
In even dimensions, we adopt the convenient convention $a_{N+1} = 0$. The coordinates $\mu_i$ are subject to the constraint
\be
\sum_{i=1}^{N+\epsilon} \mu_i^2 = 1 \, .
\ee
The spacetime generically possesses $N$ independent angular momenta $J_i$, parameterized by $N$ rotation parameters $a_i$.

The thermodynamic quantities are \cite{JHEP0615096}
\be\ba
M &= \frac{m\mathcal{A}_{d-2}}{4\pi\prod_{k\neq j}\Xi_k}\left(\sum_{i\neq j}\frac{1}{\Xi_i} +\frac{1 +\epsilon}{2} \right) \, , \\
\Omega_j &= \frac{l}{\rho_h^2} \, , \quad \Omega_{i\neq j} = \frac{\revise{a_i} (l^2 +r_h^2)}{l^2(r_h^2 +a_i^2)} \, ,  \\
J_j &= \frac{l m \mathcal{A}_{d-2}}{4\pi\prod_{k\neq j}\Xi_k} \, , \quad J_{i\neq j} = \frac{a_i m \mathcal{A}_{d-2}}{4\pi\Xi_i\prod_{k\neq j}\Xi_k} \, ,  \\
T &= \frac{1}{2\pi}\left[\frac{r_h}{l^2}\left(1 +\sum_{i\neq j}^N\frac{\rho_h^2}{r_h^2 +a_i^2}\right) -\frac{1}{r_h}\left(\frac{1}{2} -\frac{r_h^2}{2l^2} \right)^\epsilon\right] \, , \\
A &= \frac{\mathcal{A}_{d-2}}{r_h^{1-\epsilon}}\rho_h^2\prod_{i\neq j}^N\frac{r_h^2 +a_i^2}{\Xi_i} \, , \qquad S = \frac{A}{4} \, , \\
V &= \frac{r_hA}{d-1} +\frac{8\pi}{(d-1)(d-2)}\sum_{i\neq j}a_iJ_i \, .
\ea\ee
The horizon radius $r_h$ is determined by the largest root of $F -2m = 0$, while the area of the unit $(d-2)$-sphere is given by $\mathcal{A}_{d-2} = 2\pi^{[(d-1)/2]}/\Gamma[(d-1)/2]$.

We now turn to a detailed examination of the thermodynamic topological classification of ultraspinning Kerr-AdS black holes. To illustrate the general features, we begin with the four-dimensional and five-dimensional cases as representative examples of even- and odd-dimensional configurations, respectively. Each case is discussed in a separate subsection, and the universal results for higher-dimensional configurations ($d \ge 6$) are summarized in Sec. \ref{IIIC}.

%%%%%%%%%%%%%%%%%%%%%%%%%%
\subsection{$d = 4$ case}
%%%%%%%%%%%%%%%%%%%%%%%%%%

%%%%%%%%%%%%%%%%%%%%%%%%%%%%%%%%%%%%%%%%%%%%%%%%%%%%%%%%%%%%%%%%%%%%%%%%%%%%%%%%%%%%%%%%%%
\begin{figure}[t]
\centering
\includegraphics[width=0.35\textwidth]{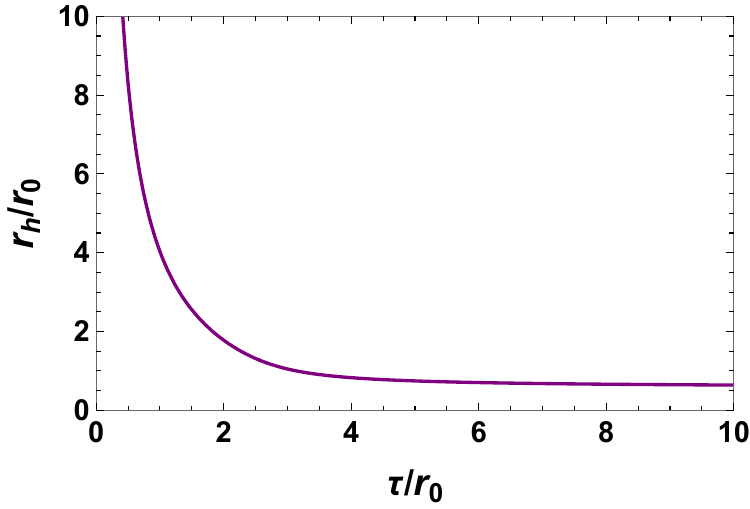}
\caption{Zero points of the vector $\phi^{r_h}$ on the $r_h$-$\tau$ plane for the four-dimensional ultraspinning Kerr-AdS black hole with $l = r_0$. The purple line denotes the thermodynamically stable branch ($w = 1$), corresponding to a topological number $W = 1$.
\label{fig1}}
\end{figure}
%%%%%%%%%%%%%%%%%%%%%%%%%%%%%%%%%%%%%%%%%%%%%%%%%%%%%%%%%%%%%%%%%%%%%%%%%%%%%%%%%%%%%%%%%%

The four-dimensional case serves as a concrete example of an even-dimensional configuration. It allows us to examine in detail the topological (sub)class structure and the asymptotic thermodynamic behavior, providing a foundation for comparison with odd-dimensional cases.

At the asymptotic boundaries, the Hawking temperature of the four-dimensional ultraspinning Kerr-AdS black hole approaches zero as $r \to r_m$, with $r_m$ denoting the minimal horizon radius, and diverges in the limit $r \to \infty$. This implies that the inverse temperature $\tau$ behaves as
\be
\tau(r_m) = \infty \, , \qquad \tau(\infty) = 0 \, .
\ee
Substituting the relation $\mathcal{A}_2 = 4\pi$ into the definition of the generalized off-shell Helmholtz free energy in Eq. (\ref{FE}) simplifies the expression for $\mathcal{F}$, yielding
\be
\mathcal{F} = \frac{(r_h^2 +l^2)^2}{2l^2r_h} -\frac{\pi(r_h^2 +l^2)}{\tau} \, ,
\ee
Accordingly, the components of the vector $\phi$ read
\bea
\phi^{r_h} &=& \frac{(3r_h^2 -l^2)(r_h^2 +l^2)}{2l^2r_h^2} -\frac{2\pi r_h}{\tau} \\
\phi^{\Theta} &=& -\cot\Theta\csc\Theta \, .
\eea
The zero point of the vector field $\phi^{r_h}$ is therefore given by
\be
\tau = \frac{4\pi l^2r_h^3}{3r_h^4 +2r_h^2l^2 -l^4} \, .
\ee

The topological number for the four-dimensional ultraspinning Kerr-AdS black hole (with $l = r_0$, where $r_0$ is the cavity length scale) is found to be $W = 1$. This result is derived from the zero points of $\phi^{r_h}$ shown in Fig. \ref{fig1} and corresponds to a thermodynamic phase structure consisting solely of a stable branch ($w = +1$).

The four-dimensional ultraspinning Kerr-AdS black hole admits a structured systematic ordering of states governed by thermodynamic topology. At least one state with positive heat capacity and winding number $w = +1$ is required. Since the heat capacity alternates in sign as the horizon radius $r_h$ increases, any further states must appear in alternating pairs. This implies that both the smallest and largest states in the $r_h$-ordered sequence are thermodynamically stable. Accordingly, the winding numbers at all zero points obey $[+, (-,+), \ldots, +]$, with the ellipsis denoting repeated alternating pairs. The full topological classification follows solely from the signs of the innermost and outermost winding numbers.

Then, we examine the universal thermodynamic behavior of the four-dimensional ultraspinning Kerr-AdS black hole. The low-temperature limit $\tau \to \infty$ is characterized by a stable small black hole state, whereas the high-temperature limit $\tau \to 0$ is characterized by a stable large black hole state.

Within the thermodynamic topological classification framework of Ref. \cite{PRD110-L081501}, the four-dimensional ultraspinning Kerr-AdS black hole is identified as a member of the class $W^{1+}$.

%%%%%%%%%%%%%%%%%%%%%%%%%%
\subsection{$d = 5$ case}
%%%%%%%%%%%%%%%%%%%%%%%%%%

%%%%%%%%%%%%%%%%%%%%%%%%%%%%%%%%%%%%%%%%%%%%%%%%%%%%%%%%%%%%%%%%%%%%%%%%%%%%%%
\begin{figure}[htbp]
\subfigure[]
{\label{fig2a}
\includegraphics[width=0.35\textwidth]{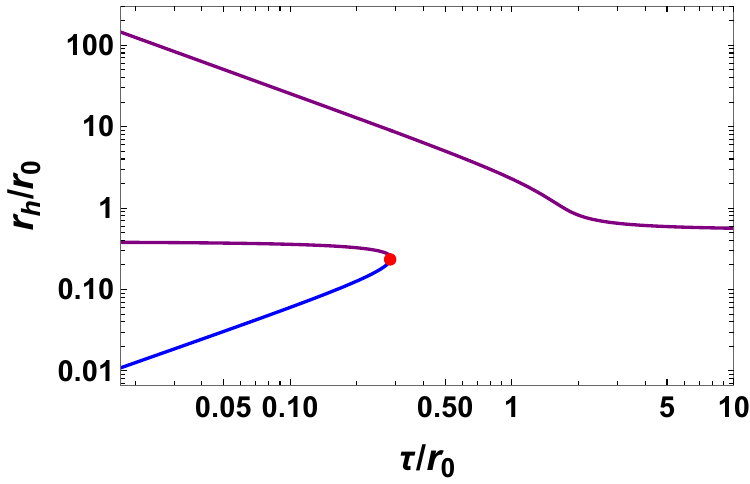}}
\subfigure[]
{\label{fig2b}
\includegraphics[width=0.35\textwidth]{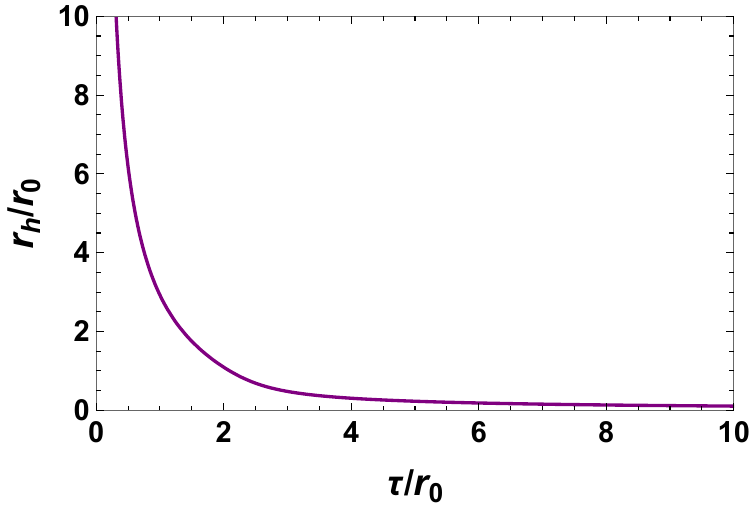}}
\caption{Zero points of $\phi^{r_h}$ in the $\tau$-$r_h$ plane for the five-dimensional ultraspinning Kerr-AdS black hole with $l = r_0$: (a) doubly rotating case with $a_1/r_0 = 0.5$; (b) singly rotating case with $a_1 = 0$. The blue curve represents a thermodynamically unstable black hole branch with winding number $w = -1$, while the purple curve corresponds to a thermodynamically stable branch with winding number $w = +1$. The red dot denotes the annihilation point, whose position in panel (a) is determined by $d\tau/dr_h=0$.} \label{fig2}
\end{figure}
%%%%%%%%%%%%%%%%%%%%%%%%%%%%%%%%%%%%%%%%%%%%%%%%%%%%%%%%%%%%%%%%%%%%%%%%%%%%%%

In this subsection, we turn to discuss the five-dimensional case, representing the odd-dimensional configurations. This example highlights the role of the maximal number of rotation parameters in determining the thermodynamic topological (sub)class, and it allows for a direct comparison of asymptotic behavior with the even-dimensional case discussed previously. In the terminology used below, the ``singly rotating" case refers to the ultraspinning solution with no additional ordinary rotation, namely $a_1=0$ besides the ultraspinning direction, while the ``doubly rotating" case refers to the ultraspinning solution with one nonzero ordinary rotation parameter, $a_1\neq0$.

After substituting $\mathcal{A}_3 = 2\pi^2$ into Eq.~(\ref{FE}), the generalized off-shell Helmholtz free energy reduces to
\bea
\mathcal{F} &=& \frac{\pi(r_h^2 +l^2)^2(r_h^2 +a_1^2)(2 +\Xi_1)}{8l^2\Xi_1^2r_h^2} \nn \\
&&-\frac{\pi^2(r_h^2 +l^2)(r_h^2 +a_1^2)}{2\tau\Xi_1r_h} \, .
\eea
The associated vector field $\phi$ is then explicitly given by
\bea
\phi^{r_h} &=& \frac{\pi(r_h^2 +l^2)[2r_h^4 +a_1^2(r_h^2 -l^2)](2 +\Xi_1)}{4l^2\Xi_1^2r_h^3} \nn \\
&&-\frac{\pi^2[3r_h^4 +r_h^2(l^2 +a_1^2) -l^2a_1^2]}{2\tau\Xi_1r_h^2} \, , \\
\phi^{\Theta} &=& -\cot\Theta\csc\Theta \, .
\eea
The zero points of the vector field are determined by solving $\phi^{r_h}=0$, which leads to
\be
\tau = \frac{2\pi l^2\Xi_1r_h[3r_h^4 +r_h^2(l^2 +a_1^2) -l^2a_1^2]}{(r_h^2 +l^2)[2r_h^4 +a_1^2(r_h^2 -l^2)](2 +\Xi_1)} \, ,
\ee
thereby fixing the inverse temperature $\tau$ at the defect points of the thermodynamic parameter space. For the doubly rotating example plotted in Fig.~\ref{fig2}(a), the annihilation point is obtained from $d\tau/dr_h=0$ together with Eq.~(19). For $l=r_0$ and $a_1/r_0=0.5$, this gives $r_h/r_0\simeq 0.2428$ and $\tau/r_0\simeq 0.2855$.

For the choice $l = r_0$, Fig. \ref{fig2} shows the zero points of $\phi^{r_h}$ in the $\tau$-$r_h$ plane for the doubly-rotating case ($a_1/r_0 = 0.5$) and the singly-rotating case ($a_1 = 0$), respectively. Both configurations yield the same global topological number $W = 1$.

As shown in Fig. \ref{fig2}, the inverse temperature $\tau$ exhibits distinct asymptotic structures for the two rotational configurations. In the doubly-rotating case, $\tau$ vanishes both at the minimal horizon radius $r_m$ and in the asymptotic regime $r_h \to \infty$. In the singly-rotating case, however, $\tau$ diverges at $r_h = r_m$ and tends to zero as $r_h \to \infty$, where $r_m$ denotes the minimal event horizon radius. The corresponding asymptotic limits are
\bea
{\rm{doubly-rotating~case}} &:& \tau(r_m) = 0 \, ,\qquad \tau(\infty) = 0 \, , \quad \\
{\rm{singly-rotating~case}} &:& \tau(r_m) = \infty \, ,\quad~~ \tau(\infty) = 0 \, . \quad
\eea

We next consider the systematic ordering of the zero points for the two cases.
In the doubly-rotating configuration, the spectrum contains at least three black hole states: two stable states with $w = +1$ and one unstable state with $w = -1$. Additional states can only appear in pairs with opposite winding numbers. The winding number sequence is therefore nontrivial. It begins with a pair $[-,+]$, followed by a block structure that starts with $+$, ends with $+$, and contains alternating $[-,+]$ pairs in between, giving rise to a subsequence of the form $[+ ,- ,+ ,-, \ldots, +]$ This implies that the smallest black hole is unstable, whereas the largest one is stable, so that the topological classification is fixed by the endpoint configuration $[-,+]$. For the singly-rotating case, however, the sequence follows the pattern
$[+, (-,+), \ldots, (-,+)]$, implying that both the smallest and largest black holes are thermodynamically stable.

Then, we analyze the asymptotic thermodynamic behavior in the two configurations. In the low-temperature limit $(\tau \to \infty)$, both the doubly-rotating and singly-rotating cases admit only a single thermodynamically stable small black hole state. In the high-temperature limit $(\tau \to 0)$, the phase structures differ qualitatively. The doubly-rotating configuration exhibits a richer structure, consisting of three distinct branches: a stable large black hole, a stable small black hole, and an unstable small black hole. In contrast, the singly-rotating configuration admits only a single stable large black hole branch.

In conclusion, our analysis establishes a clear topological distinction for five-dimensional ultraspinning Kerr-AdS black holes. Within the thermodynamic topological classification framework of Refs. \cite{PRD110-L081501,
PRD112-124024}, the singly-rotating configuration falls into the known topological class $W^{1+}$, whereas the doubly-rotating configuration is assigned to the distinct subclass $\tilde{W}^{1+}$.

%%%%%%%%%%%%%%%%%%%%%%%%%%%%%%%%%%%%%%%%
\subsection{$d \ge 6$ cases}\label{IIIC}
%%%%%%%%%%%%%%%%%%%%%%%%%%%%%%%%%%%%%%%%

\onecolumngrid

\begin{center}
\begin{table}[h]
\caption{Thermodynamical properties of the black hole states for the ultraspinning Kerr-AdS black holes in arbitrary dimensions. The abbreviations denote degenerate point (DP) and annihilation point (AP), respectively.}\label{TableI}
\begin{tabular}{c|c|c|c|c|c|c|c}
\hline\hline
Cases & Classes  & Innermost & Outermost & \makecell{Low $T$ \\ ($\tau\to\infty$)} & \makecell{High $T$ \\ ($\tau\to 0$)} & DP/AP & $W$ \\ \hline
Even-dimensional ultraspinning Kerr-AdS black holes & $W^{1+}$  & stable  & stable & stable  small & stable  large & in pairs & $+1$ \\ \hline
\makecell{Odd-dimensional ultraspinning Kerr-AdS \\ black holes with fewer rotations} & $W^{1+}$  & stable  & stable & stable small & stable  large & in pairs & $+1$  \\ \hline
\makecell{Odd-dimensional ultraspinning Kerr-AdS \\ black holes with maximal rotations} & $\tilde{W}^{1+}$  & unstable & stable & stable small & \makecell{unstable small + \\ stable small + \\ stable large} & one more AP & $+1$ \\
\hline\hline
\end{tabular}
\end{table}
\end{center}

\twocolumngrid

In Table~\ref{TableI}, DP and AP are identified from the zero-point curve $\tau(r_h)$. A degenerate point (DP) denotes a local degeneracy where neighboring zero points merge, equivalently where $\phi^{r_h}=0$ and $\partial\phi^{r_h}/\partial r_h=0$, or $d\tau/dr_h=0$ on the on-shell curve. When the curve is followed as the control parameter $\tau$ crosses the critical value, such a degeneracy is classified as a generation or annihilation event according to whether the neighboring zero points appear or disappear; AP denotes the annihilation type. For the $W^{1+}$ cases these DP/AP events appear in compensating pairs and do not change the stable endpoint pattern. In the odd-dimensional maximal-rotation sector, the small-radius endpoint already starts from an unstable branch, and the corresponding zero-point sequence contains one additional AP, giving the subclass $\tilde W^{1+}$.

The analysis for four- and five-dimensional configurations can be systematically extended to ultraspinning Kerr-AdS black holes in arbitrary higher dimensions ($d \ge 6$). Following the same thermodynamic topological procedure outlined in Sec. \ref{II}, we have explicitly constructed the off-shell free energy and vector field $\phi$, and numerically analyzed the zero-point structures for representative cases with $d = 6, 7, \ldots, 11$, including both maximal-rotation and non-maximal-rotation configurations. No new topological structures arise beyond those identified in lower dimensions.

The reason that these representative cases support the general classification is that the topology is controlled by the endpoint behavior of the zero-point curve $\tau(r_h)=1/T$. From Eq.~(10), the large-radius behavior is $T\simeq (d-1)r_h/(4\pi l^2)$, and hence $\tau\simeq 4\pi l^2/[(d-1)r_h]\to0$ as $r_h\to\infty$, corresponding to a stable large-black-hole branch. Zero points can be created or annihilated only at turning points of this curve; within a fixed endpoint pattern these local events occur as opposite-winding pairs and therefore do not generate a new topological class or subclass. The small-radius endpoint is controlled by the lowest nonvanishing term in $\phi^{r_h}$, which changes according to whether an ordinary rotation direction is absent. In even dimensions the convention $a_{N+1}=0$, and in odd dimensions with fewer than maximal rotations at least one ordinary rotation parameter is absent; in both cases the zero-point sequence starts from a stable small-black-hole branch. By contrast, odd-dimensional maximal-rotation configurations are the only cases in which no ordinary rotation parameter is forced or chosen to vanish after the ultraspinning limit. The leading small-radius behavior then changes so that the innermost branch is unstable, reproducing the endpoint pattern of the five-dimensional doubly rotating case.

In the explicit five-dimensional ultraspinning examples used to calibrate the numerical analysis, we set $l=r_0=1$ and used $a_1/r_0=0.5$ for the doubly rotating case, while the singly rotating case corresponds to $a_1=0$. For the higher-dimensional checks with $d=6,\ldots,11$, the scan was organized by rotation pattern rather than by a single tuned parameter set: for each dimension we compared maximal-rotation sectors, with all ordinary rotation parameters nonzero in the under-rotating regime, against non-maximal sectors obtained by setting one or more ordinary $a_i$ to zero. We then varied the nonzero $a_i$ within the physical range $0<a_i<l$ and monitored the endpoint winding sequence and the DP/AP structure. No additional endpoint pattern was found in these checks, and hence no additional topological class or subclass appears.

The key findings are as follows:
\begin{enumerate}
\item
Even dimensions ($d = 6, 8, ...$): All configurations, regardless of the number of independent rotation parameters, exhibit a topological structure identical to that of the four-dimensional case. The zero-point sequence begins and ends with stable branches ($w = +1$), and the global topological number is consistently $W = 1$. Consequently, all even-dimensional ultraspinning Kerr-AdS black holes belong to the topological class $W^{1+}$.

\item
Odd dimensions ($d = 7, 9, ...$): The classification bifurcates according to the number of rotations. Configurations with the maximal number of independent rotation parameters display the same endpoint pattern as the five-dimensional doubly-rotating black hole: the innermost zero point is unstable ($w = -1$), while the outermost is stable ($w = +1$). These belong to the distinct subclass
$\tilde{W}^{1+}$. In contrast, odd-dimensional black holes with fewer than the maximal rotations follow the pattern of the five-dimensional singly-rotating case, where both endpoints are stable, and are therefore classified as $W^{1+}$.
\end{enumerate}

These higher-dimensional results support and generalize the unified classification rule valid across dimensions summarized in Table \ref{TableI}. The thermodynamic topology of ultraspinning Kerr-AdS black holes is governed jointly by spacetime dimensionality and the attainment of the maximal rotational structure. Within the endpoint analysis and representative higher-dimensional checks, no new topological classes or subclasses beyond $W^{1+}$ and $\tilde{W}^{1+}$ were found, supporting the proposed scheme.

%%%%%%%%%%%%%%%%%%%%%%%%%%%%%%%
\section{Conclusions}\label{IV}
%%%%%%%%%%%%%%%%%%%%%%%%%%%%%%%
We have systematically classified ultraspinning Kerr-AdS black holes in arbitrary dimensions using the thermodynamic topology framework, in which black holes are treated as topological defects in the off-shell thermodynamic parameter space. Their phase structures are characterized by global invariants, including zero points, winding numbers, and asymptotic thermodynamic behavior. By analyzing the ordering of stable and unstable branches, we obtain a unified classification scheme that applies across all dimensions.

The four-dimensional and five-dimensional cases were examined in detail as representative examples of even- and odd-dimensional configurations, and the corresponding higher-dimensional cases ($d \ge 6$) were systematically analyzed within the same framework, as summarized in Sec. \ref{IIIC}. These analyses consistently indicate that ultraspinning Kerr-AdS black holes possess topological structures that remain invariant under local deformations preserving the endpoint pattern and reflect the global organization of thermodynamic states.

Our main result is a simple classification rule valid across spacetime dimensions. In odd spacetime dimensions, ultraspinning Kerr-AdS black holes with the maximal number of independent rotation parameters belong to the distinct subclass $\tilde{W}^{1+}$. All other configurations, including odd-dimensional black holes with fewer rotations and all even-dimensional ultraspinning black holes, belong to the standard topological class $W^{1+}$. This classification is determined by the ordering of stable and unstable states and the winding numbers at the endpoints of the zero-point sequences. Maximal-rotation configurations in odd dimensions exhibit a qualitatively different topological organization, while all other configurations follow the standard pattern. These classification results and the corresponding thermodynamical properties of the black hole states are summarized in Table \ref{TableI} for clarity. Heuristically, the appearance of the subclass $\tilde{W}^{1+}$ in odd dimensions with maximal rotations suggests that the full saturation of rotational degrees of freedom plays a nontrivial role in shaping the global thermodynamic topology of the phase space.

This result also clarifies its relation to ordinary Kerr-AdS black holes. In the earlier thermodynamic-topology analysis of singly rotating Kerr-AdS black holes \cite{PRD107-084002}, the results fall into the standard classes of the asymptotic classification scheme \cite{PRD110-L081501}: the four- and five-dimensional cases belong to $W^{1+}$, whereas the singly rotating cases with $d\geq 6$ belong to $W^{0-}$. The later study of multiply rotating Kerr-AdS black holes identified the subclass $\tilde{W}^{1+}$ in the odd-dimensional sectors carrying the maximal number of independent rotation parameters \cite{PRD112-124024}. Together with the corresponding non-maximal multiply rotating cases, which follow the same standard patterns as the singly rotating classification, this comparison indicates that $\tilde{W}^{1+}$ is tied to the simultaneous conditions of odd spacetime dimension and maximal rotation, rather than merely to the presence of multiple rotation parameters. The present ultraspinning analysis shows that this parity-and-maximal-rotation criterion persists after taking the ultraspinning limit: only odd-dimensional maximal-rotation ultraspinning Kerr-AdS black holes enter $\tilde{W}^{1+}$, while the remaining ultraspinning sectors stay in $W^{1+}$.

These findings provide a unified topological interpretation of ultraspinning Kerr-AdS black holes and a classification framework valid across dimensions for understanding their thermodynamic phase structures. The approach can be extended to charged ultraspinning solutions \cite{JHEP0615096,PRD95-046002}, supergravity embeddings\cite{PRD102-044007,PRD103-044014,JHEP1121031,2501.13655}, and spacetimes with nontrivial asymptotics \cite{PRD105-024013,PLB846-138227,PRD108-064034,PRD108-064035,2606.03958,JCAP0924017,
PRD110-104072,2601.14315}.

\acknowledgments

We are very grateful to the anonymous referee for valuable comments and suggestions.
This work is supported by the National Natural Science Foundation of China (NSFC) under Grants No. 12205243, No. 12375053, by the Sichuan Science and Technology Program under Grant No. 2026NSFSC0021, and by the Doctoral Research Initiation Project of China West Normal University under Grant No. 23KE026.


\begin{thebibliography}{99}
%\def\CQG{Class. Quant. Grav.\,}
\def\CQG{Classical Quantum Gravity\,}
\def\EPJC{Eur. Phys. J. C\,}
%\def\GRG{Gen. Relativ. Gravit.\,}
\def\GRG{Gen. Rel. Grav.\,}
\def\JCAP{J. Cosmol. Astropart. Phys.\,}
\def\JHEP{J. High Energy Phys.\,}
\def\PRD{Phys. Rev. D\,}
\def\PDU{Phys. Dark Univ.\,}
\def\PRL{Phys. Rev. Lett.\,}
\def\NPB{Nucl. Phys. B \,}
\def\PLB{Phys. Lett. B \,}
\def\JMP{J. Math. Phys. (N.Y.)\,}
\def\MPLA{Mod. Phys. Lett. A\,}
\def\AP{Ann. Phys. (N.Y.)\,}
\def\APJ{Astrophys. J.\,}
\def\APJL{Astrophys. J. Lett.\,}
\def\CPL{Chin. Phy. Lett.\,}
\def\IJMPA{Int. J. Mod. Phys. A\,}
\def\PRR{Phys. Rev. Res.\,}

\bibitem{PRL115-031101}
R.A. Hennigar, R.B. Mann, and D. Kubiz\v{n}\'{a}k,
Entropy Inequality Violations from Ultraspinning Black Holes,
\href{http://dx.doi.org/10.1103/PhysRevLett.115.031101}
{\PRL \textbf{115}, 031101 (2015)}.

\bibitem{JHEP0114127}
A. Gnecchi, K. Hristov, D. Klemm, C. Toldo, and O. Vaughan,
Rotating black holes in 4d gauged supergravity,
\href{http://dx.doi.org/10.1007/JHEP01(2014)127}
{\JHEP \textbf{01} (2014) 127}.

\bibitem{PRD89-084007}
D. Klemm,
Four-dimensional black holes with unusual horizons,
\href{http://dx.doi.org/10.1103/PhysRevD.89.084007}
{\PRD \textbf{89}, 084007 (2014)}.

\bibitem{PRD84-024037}
M. Cveti\v{c}, G.W. Gibbons, D. Kubiz\v{n}\'{a}k, and C.N. Pope,
Black hole enthalpy and an entropy inequality for the thermodynamic volume,
\href{http://dx.doi.org/10.1103/PhysRevD.84.024037}
{\PRD \textbf{84}, 024037 (2011)}.

\bibitem{PRD87-104017}
B.P. Dolan, D. Kastor, D. Kubiz\v{n}\'{a}k, R.B. Mann, and J. Traschen,
Thermodynamic volumes and isoperimetric inequalities for de Sitter black holes,
\href{http://dx.doi.org/10.1103/PhysRevD.87.104017}
{\PRD \textbf{87}, 104017 (2013)}.

\bibitem{JHEP0615096}
R.A. Hennigar, D. Kubiz\v{n}\'{a}k, R.B. Mann, and N. Musoke,
Ultraspinning limits and super-entropic black holes,
\href{http://dx.doi.org/10.1007/JHEP06(2015)096}
{\JHEP \textbf{06} (2015) 096}.

\bibitem{PRD95-046002}
S.M. Noorbakhsh and M. Ghominejad,
Ultra-spinning gauged supergravity black holes and their Kerr/CFT correspondence,
\href{http://dx.doi.org/10.1103/PhysRevD.95.064002}
{\PRD \textbf{95}, 046002 (2017)}.

\bibitem{1702.03448}
S.M. Noorbakhsh and M. Ghominejad,
Higher dimensional charged AdS black holes at ultra-spinning limit and their 2d CFT duals,
\href{https://arxiv.org/abs/1702.03448}{arXiv:1702.03448}.

\bibitem{JHEP0118042}
S.M. Noorbakhsh and M.H. Vahidinia,
Extremal vanishing horizon Kerr-AdS black holes at ultraspinning limit,
\href{http://dx.doi.org/10.1007/JHEP01(2018)042}
{\JHEP \textbf{01} (2018) 042}.

\bibitem{PRD102-044007}
D. Wu, P. Wu, H. Yu, and S.-Q. Wu,
Are ultraspinning Kerr-Sen-AdS$_4$ black holes always superentropic?,
\href{http://dx.doi.org/10.1103/PhysRevD.102.044007}
{\PRD \textbf{102}, 044007 (2020)}.

\bibitem{PRD103-044014}
D. Wu, S.-Q. Wu, P. Wu, and H. Yu,
Aspects of the dyonic Kerr-Sen-AdS$_4$ black hole and its ultraspinning version,
\href{http://dx.doi.org/10.1103/PhysRevD.103.044014}
{\PRD \textbf{103}, 044014 (2021)}.

\bibitem{JHEP1121031}
D. Wu and S.-Q. Wu,
Ultra-spinning Chow's black holes in six-dimensional gauged supergravity and their
thermodynamical properties,
\href{http://dx.doi.org/10.1007/JHEP11(2021)031}
{\JHEP \textbf{11} (2021) 031}.

\bibitem{JHEP0220195}
M. Appels, L. Cuspinera, R. Gregory, P. Krtou\v{s}, and D. Kubiz\v{n}\'{a}k,
Are superentropic black holes superentropic?,
\href{http://dx.doi.org/10.1007/JHEP02(2020)195}
{\JHEP \textbf{02} (2020) 195}.

\bibitem{MPLA35-2050098}
C.V. Johnson,
Instability of super-entropic black holes in extended thermodynamics,
\href{http://dx.doi.org/10.1142/S0217732320500984}
{\MPLA \textbf{35}, 2050098 (2020)}.

\bibitem{PRD101-086006}
C.V. Johnson, V.L. Martin, and A. Svesko,
A microscopic description of thermodynamic volume in extended black hole thermodynamics,
\href{http://dx.doi.org/10.1103/PhysRevD.101.086006}
{\PRD \textbf{101}, 086006 (2020)}.

\bibitem{PRD101-024057}
D. Wu, P. Wu, H. Yu, and S.-Q. Wu,
Notes on the thermodynamics of superentropic AdS black holes,
\href{http://dx.doi.org/10.1103/PhysRevD.101.024057}
{\PRD \textbf{101}, 024057 (2020)}.

\bibitem{PLB807-135529}
Z.M. Xu,
The correspondence between thermodynamic curvature and isoperimetric theorem from
ultraspinning black hole,
\href{http://dx.doi.org/10.1016/j.physletb.2020.135529}
{\PLB \textbf{807}, 135529 (2020)}.

\bibitem{2601.22565}
Z. Di,
Thermodynamics and stability of ultraspinning black holes,
\href{https://arxiv.org/abs/2601.22565}{arXiv:2601.22565}.

\bibitem{JHEP0816148}
M. Sinamuli and R.B. Mann,
Super-entropic black holes and the Kerr-CFT correspondence,
\href{http://dx.doi.org/10.1007/JHEP08(2016)148}
{\JHEP \textbf{08} (2016) 148}.

\bibitem{1912.03974}
K. Flathmann and N. Wassermann,
Geodesic equations for particles and light in the black spindle spacetime,
\href{http://doi.org/10.1063/5.0011432}
{\JMP \textbf{61}, 122504 (2020)}.

\bibitem{CQG38-045018}
M.T.N. Imseis, A.A Balushi, and R.B. Mann,
Null hypersurfaces in Kerr-Newman-AdS black hole and super-entropic black hole spacetimes,
\href{http://dx.doi.org/10.1088/1361-6382/abd3e0}
{\CQG \textbf{38}, 045018 (2021)}.

\bibitem{PRD103-024053}
S. Noda and Y.C. Ong,
Null hypersurface caustics, closed null curves, and super entropy,
\href{http://dx.doi.org/10.1103/PhysRevD.103.024053}
{\PRD \textbf{103}, 024053 (2021)}.

\bibitem{PRD103-104020}
D. Wu and P. Wu,
Null hypersurface caustics for high-dimensional superentropic black holes,
\href{http://dx.doi.org/10.1103/PhysRevD.103.104020}
{\PRD \textbf{103}, 104020 (2021)}.

\bibitem{PRD104-L121501}
D. Wu, P. Wu, and H. Yu,
Shadowless rapidly rotating yet not ultraspinning Kerr-AdS$_4$ and Kerr-Newman-AdS$_4$ black holes,
\href{http://dx.doi.org/10.1103/PhysRevD.104.L121501}
{\PRD \textbf{104}, L121501 (2021)}.

\bibitem{PLB821-136619}
A. Belhaj, H. Belmahi, and M. Benali,
Superentropic AdS black hole shadows,
\href{https://doi.org/10.1016/j.physletb.2021.136619}
{\PLB \textbf{821}, 136619 (2021)}.

\bibitem{EPJC82-619}
A. Belhaj, M. Benali, and Y. Hassouni,
Superentropic black hole shadows in arbitrary dimensions,
\href{https://doi.org/10.1140/epjc/s10052-022-10564-x}
{\EPJC \textbf{82}, 619 (2022)}.

\bibitem{PRL129-191101}
S.-W. Wei, Y.-X. Liu, and R.B. Mann,
Black Hole Solutions as Topological Thermodynamic Defects,
\href{https://doi.org/10.1103/PhysRevLett.129.191101}
{\PRL \textbf{129}, 191101 (2022)}.

\bibitem{PRD110-L081501}
S.-W. Wei, Y.-X. Liu, and R.B. Mann,
Universal topological classifications of black hole thermodynamics,
\href{https://doi.org/10.1103/PhysRevD.110.L081501}
{\PRD \textbf{110}, L081501 (2024)}.

\bibitem{PRD111-L061501}
D. Wu, W. Liu, S.-Q. Wu, and R.B. Mann,
Novel topological classes in black hole thermodynamics,
\href{https://doi.org/10.1103/PhysRevD.111.L061501}
{\PRD \textbf{111}, L061501 (2025)}.

\bibitem{PRD112-124024}
W. Ai and D. Wu,
$\tilde{W}^{1+}$: Extending the topological classification of black hole thermodynamics,
\href{https://doi.org/10.1103/dy7y-j24r}
{\PRD \textbf{112}, 124024 (2025)}.

\bibitem{2510.20164}
D. Wu and S.-Q. Wu,
Thermodynamics and topological classifications of static non-extremal four-charge AdS black hole in the five-dimensional
$\mathcal{N}=2$, $STU-W^2U$ gauged supergravity,
\href{https://arxiv.org/abs/2510.20164}{arXiv:2510.20164}.

\bibitem{EPJC85-1386}
H. Chen, M.-Y. Zhang, H. Hassanabadi, Q. Huang, Z.-W. Long,
Novel topological subclass in Ho\v{r}ava-Lifshitz black holes,
\href{https://doi.org/10.1140/epjc/s10052-025-15133-6}
{\EPJC \textbf{85}, 1386 (2025)}.

\bibitem{PRD107-024024}
D. Wu,
Topological classes of rotating black holes,
\href{https://doi.org/10.1103/PhysRevD.107.024024}
{\PRD \textbf{107}, 024024 (2023)}.

\bibitem{PRD107-064023}
C.H. Liu and J. Wang,
The topological natures of the Gauss-Bonnet black hole in AdS space,
\href{https://doi.org/10.1103/PhysRevD.107.064023}
{\PRD \textbf{107}, 064023 (2023)}.

\bibitem{PRD107-084002}
D. Wu and S.-Q. Wu,
Topological classes of thermodynamics of rotating AdS black holes,
\href{https://doi.org/10.1103/PhysRevD.107.084002}
{\PRD \textbf{107}, 084002 (2023)}.

\bibitem{EPJC83-365}
D. Wu,
Classifying topology of consistent thermodynamics of the four-dimensional
neutral Lorentzian NUT-charged spacetimes,
\href{https://doi.org/10.1140/epjc/s10052-023-11561-4}
{\EPJC \textbf{83}, 365 (2023)}.

\bibitem{EPJC83-589}
D. Wu,
Consistent thermodynamics and topological classes for the four-dimensional
Lorentzian charged Taub-NUT spacetimes,
\href{https://doi.org/10.1140/epjc/s10052-023-11782-7}
{\EPJC \textbf{83}, 589 (2023)}.

\bibitem{PRD108-084041}
D. Wu,
Topological classes of thermodynamics of the four-dimensional static accelerating black holes,
\href{https://doi.org/10.1103/PhysRevD.108.084041}
{\PRD \textbf{108}, 084041 (2023)}.

\bibitem{JHEP0624213}
D. Wu, S.-Y. Gu, X.-D. Zhu, Q.-Q. Jiang, and S.-Z. Yang,
Topological classes of thermodynamics of the static multi-charge AdS black holes in gauged supergravities: novel temperature-dependent thermodynamic topological phase transition,
\href{https://doi.org/10.1007/JHEP06(2024)213}
{\JHEP \textbf{06} (2024) 213}.

\bibitem{PLB856-138919}
X.-D. Zhu, D. Wu, and D. Wen,
Topological classes of thermodynamics of the rotating charged AdS black holes in gauged supergravities,
\href{https://doi.org/10.1016/j.physletb.2024.138919}
{\PLB \textbf{856}, 138919 (2024)}.

\bibitem{PDU46-101617}
H. Chen, D. Wu, M.-Y. Zhang, H. Hassanabadi, and Z.-W. Long,
Thermodynamic topology of phantom AdS black holes in massive gravity,
\href{https://doi.org/10.1016/j.dark.2024.101617}
{\PDU \textbf{46}, 101617 (2024)}.

\bibitem{EPJC84-1294}
Z.-Q. Chen, and S.-W. Wei,
Thermodynamical topology with multiple defect curves for dyonic AdS black holes,
\href{https://doi.org/10.1140/epjc/s10052-024-13620-w}
{\EPJC \textbf{84}, 1294 (2024)}.

\bibitem{CQG42-125007}
W. Liu, L. Zhang, D. Wu, and J. Wang,
Thermodynamic topological classes of the rotating, accelerating black holes,
\href{https://doi.org/10.1088/1361-6382/ade35b}
{\CQG \textbf{42}, 125007 (2025)}.

\bibitem{PLB860-139163}
X.-D. Zhu, W. Liu, and D. Wu,
Universal thermodynamic topological classes of rotating black holes,
\href{https://doi.org/10.1016/j.physletb.2024.139163}
{\PLB \textbf{860}, 139163 (2025)}.

\bibitem{PLB865-139482}
Y. Chen, X.-D. Zhu, and D. Wu,
Universal thermodynamic topological classes of three-dimensional BTZ black holes,
\href{https://doi.org/10.1016/j.physletb.2025.139482}
{\PLB \textbf{865}, 139482 (2025)}.

\bibitem{EPJC85-828}
H. Chen, D. Wu, M.-Y. Zhang, S. Zare, H. Hassanabadi, B.C. L\"utf\"uo\v{g}lu, and Z.-W. Long,
Universal thermodynamic topological classes of static black holes in Conformal Killing Gravity,
\href{https://doi.org/10.1140/epjc/s10052-025-14581-4}
{\EPJC \textbf{85}, 828 (2025)}.

\bibitem{PRL119-251102}
P.V.P. Cunha, E. Berti, and C.A.R. Herdeiro,
Light Ring Stability in Ultra-Compact Objects,
\href{http://dx.doi.org/10.1103/PhysRevLett.119.251102}
{\PRL \textbf{119}, 251102 (2017)}.

\bibitem{PRL124-181101}
P.V.P. Cunha, and C.A.R. Herdeiro,
Stationary Black Holes and Light Rings,
\href{http://dx.doi.org/10.1103/PhysRevLett.124.181101}
{\PRL \textbf{124}, 181101 (2020)}.

\bibitem{PRD102-064039}
S.-W. Wei,
Topological charge and black hole photon spheres,
\href{https://doi.org/10.1103/PhysRevD.102.064039}
{\PRD \textbf{102}, 0604039 (2020)}.

\bibitem{PRD103-104031}
M. Guo and S. Gao,
Universal properties of light rings for stationary axisymmetric spacetimes,
\href{https://doi.org/10.1103/PhysRevD.103.104031}
{\PRD \textbf{103}, 104031 (2021)}.

\bibitem{PLB858-139052}
W. Liu, D. Wu, and J. Wang,
Light rings and shadows of static black holes in effective quantum gravity,
\href{https://doi.org/10.1016/j.physletb.2024.139052}
{\PLB \textbf{858}, 139052 (2024)}.

\bibitem{PLB868-139742}
W. Liu, D. Wu, and J. Wang,
Light rings and shadows of static black holes in effective quantum gravity II: A new solution
without Cauchy horizons,
\href{https://doi.org/10.1016/j.physletb.2025.139812}
{\PLB \textbf{868}, 139742 (2025)}.

\bibitem{PRD107-064006}
S.-W. Wei and Y.-X. Liu,
Topology of equatorial timelike circular orbits around stationary black holes,
\href{https://doi.org/10.1103/PhysRevD.107.064006}
{\PRD \textbf{107}, 064006 (2023)}.

\bibitem{JCAP0723049}
X. Ye and S.-W. Wei,
Topological study of equatorial timelike circular orbit for spherically symmetric (hairy) black holes,
\href{https://doi.org/10.1088/1475-7516/2023/07/049}
{\JCAP \textbf{07} (2023) 049}.

\bibitem{PRD108-044077}
J. Yin, J. Jiang, and M. Zhang,
Kinematic topologies of black holes,
\href{https://doi.org/10.1103/PhysRevD.108.044077}
{\PRD \textbf{108}, 044077 (2023)}.

\bibitem{CQG42-025020}
X. Ye, and S.-W. Wei,
Novel topological phenomena of timelike circular orbits for charged test particles,
\href{https://doi.org/10.1088/1361-6382/ad9f14}
{\CQG \textbf{42}, 025020 (2025)}.

\bibitem{PRD15-2752}
G.W. Gibbons and S.W. Hawking,
Action integrals and partition functions in quantum gravity,
\href{https://doi.org/10.1103/PhysRevD.15.2752}
{\PRD \textbf{15}, 2752 (1977)}.

\bibitem{PRD33-2092}
J.W. York,
Black-hole thermodynamics and the Euclidean Einstein action,
\href{https://doi.org/10.1103/PhysRevD.33.2092}
{\PRD \textbf{33}, 2092 (1986)}

\bibitem{PRD105-084030}
S.-J. Yang, R. Zhou, S.W. Wei, and Y.-X. Liu,
Dynamics and kinetics of phase transition for Kerr AdS black hole on free energy landscape,
\href{https://doi.org/10.1103/PhysRevD.105.084030}
{\PRD \textbf{105}, 084030 (2022)}.

\bibitem{PRD106-106015}
R. Li and J. Wang,
Generalized free energy landscape of a black hole phase transition,
\href{https://doi.org/10.1103/PhysRevD.106.106015}
{\PRD \textbf{106}, 106015 (2022)}.

\bibitem{SS9-1072}
Y.-S. Duan and M.-L. Ge,
$SU$ (2) gauge theory and electrodynamics of $N$ moving magnetic monopoles,
\href{https://doi.org/10.1142/9789813237278_0001}
{Sci. Sin. \textbf{9}, 1072 (1979)}.

\bibitem{NPB514-705}
Y.-S. Duan, S. Li, and G.-H. Yang,
The bifurcation theory of the Gauss-Bonnet-Chern topological current and Morse function,
\href{https://doi.org/10.1016/S0550-3213(97)00777-3}
{\NPB \textbf{514}, 705 (1998)}.

\bibitem{PRD61-045004}
L.-B. Fu, Y.-S. Duan, and H. Zhang,
Evolution of the Chern-Simons vortices,
\href{https://doi.org/10.1103/PhysRevD.61.045004}
{\PRD \textbf{61}, 045004 (2000)}.

\bibitem{2501.13655}
D. Wu and S.-Q. Wu,
Four-charge static non-extremal black holes in the five-dimensional $\mathcal{N} =2$, $STU-W^2U$ supergravity,
\href{http://arxiv.org/abs/2510.13655}{arXiv:2510.13655}.

\bibitem{PRD105-024013}
D. Wu and S.-Q. Wu,
Consistent mass formulas for the four-dimensional dyonic NUT-charged spacetimes,
\href{https://doi.org/10.1103/PhysRevD.105.124013}
{\PRD \textbf{105}, 124013 (2022)}.

\bibitem{PLB846-138227}
D. Wu and S.-Q. Wu,
Revisiting mass formulas of the four-dimensional Reissner-Nordstr\"om-NUT-AdS solutions in a different metric form,
\href{https://doi.org/10.1016/j.physletb.2023.138227}
{\PLB \textbf{846}, 138227 (2023)}.

\bibitem{PRD108-064034}
D. Wu and S.-Q. Wu,
Consistent mass formulas for higher even-dimensional Taub-NUT spacetimes and their AdS counterparts,
\href{https://doi.org/10.1103/PhysRevD.108.064034}
{\PRD \textbf{108}, 064034 (2023)}.

\bibitem{PRD108-064035}
S.-Q. Wu and D. Wu,
Consistent mass formulas for higher even-dimensional Reissner-Nordstr\"{o}m-NUT-AdS spacetimes,
\href{https://doi.org/10.1103/PhysRevD.108.064035}
{\PRD \textbf{108}, 064035 (2023)}.

\bibitem{2606.03958}
D. Wu and S.-Q. Wu,
Multihair thermodynamics of Kerr-Newman-NUT-AdS$_4$ spacetimes,
\href{http://arxiv.org/abs/2606.03958}{arXiv:2606.03958}.

\bibitem{JCAP0924017}
W. Liu, D. Wu, and J. Wang,
Static neutral black holes in Kalb-Ramond gravity,
\href{https://doi.org/10.1088/1475-7516/2024/09/017}
{\JCAP \textbf{09} (2024) 017}.

\bibitem{PRD110-104072}
S.-Q. Wu and D. Wu,
Is the type-D NUT C-metric really ``missing" from the most general Pleba\'nski-Demia\'nski solution?,
\href{https://doi.org/10.1103/PhysRevD.110.104072}
{\PRD \textbf{110}, 104072 (2024)}.

\bibitem{2601.14315}
D. Wu and S.-Q. Wu,
Static four-charge squashed black hole in five-dimensional $STU-W^2U$ supergravity and its thermodynamics,
\href{http://arxiv.org/abs/2601.14315}{arXiv:2601.14315}.

\end{thebibliography}
\end{document}